\documentclass[doublecol]{manuscript}
\usepackage[mathletters]{ucs}
\usepackage[utf8]{inputenc}
\usepackage{amsmath}
\usepackage{amssymb}
\usepackage{wasysym}
\usepackage{xcolor}
\usepackage{graphicx}

\DeclareMathOperator{\sinc}{sinc}

\title{Temperature fluctuations in the Ultimate Regime of Convection}
\author{F. Gauthier\inst{1} \and J. Salort\inst{1} \and O. Bourgeois\inst{1} \and J.-L. Garden\inst{1} \and R. du Puits\inst{2} \and A. Thess\inst{2} \and P.-E. Roche\inst{1}}
\shortauthor{F. Gauthier \etal}

\institute{                    
  \inst{1} Institut N\'eel, CNRS / UJF - BP 166, F-38042 Grenoble cedex 9, France \\
  \inst{2} Technische Universität Ilmenau - Ehrenbergstraße 29, D-98693 Ilmenau, Germany
}

\abstract{
A new regime of turbulent convection has been reported nearly one decade ago, based on global heat transfer measurements at very high Rayleigh numbers. We examine the signature of this ``Ultimate Regime'' from within the flow itself. A systematic study of probe-size corrections shows that the earlier temperature measurements within the flow were altered by an excessive size of thermometer, but not according to a theoretical model proposed in the literature. Using a probe one order of magnitude smaller than the one used previously, we find evidence that the transition to the Ultimate Regime is indeed accompanied with a clear change in the statistics of temperature fluctuations in the flow.
}

\begin{document}

% cd /Users/per/Desktop/CanevasNewEvidences/Images 
% for i in *.eps; do epstopdf $i; done

\maketitle

In 1997, a new turbulent regime of thermal convection was observed by Chavanne \textit{et al.} % in a Rayleigh-B\'enard cell
 for Prandtl number of order unity ($Pr\sim 1$) and above a threshold Rayleigh number of order  $Ra\simeq 10^{12}$ (the definitions of $Pr$ and $Ra$ are recalled later)\cite{Chavanne1997}. Such conditions are found in environmental flows, including atmospheric and oceanic, giving a major practical importance to this result, beyond the theoretical challenge it raises. This new regime is characterised by an improved heat transfer, which is usually assessed by the dimensionless conductivity $Nu$ of the flow.  This Nusselt number $Nu$ is defined as the total heat flux across the cell normalised by the diffusive heat flux that would settle in a quiescent fluid. Right above the reported transition, $Nu$ scales like $Nu \sim Ra^{0.39\pm0.02}$ while the $Ra$ exponent reaches at most $1/3$ right below. A second signature of this regime was recently reported : enhanced fluctuations of the temperature drop across the boundary layer covering the plate used to force heat through the flow. This observation is consistent with an hydrodynamic boundary layer instability% boosting the shot noise of thermal plumes leaving the plates
 \cite{GauthierShotNoise:2008}. Both observations, as well as specific tests (in particular the observation of a $Nu \sim Ra^{1/2}$ heat transfer law \cite{RochePRE2001}) are fully consistent with a 1962 prediction by R. Kraichnan \cite{Kraichnan1962}. This prediction states that an asymptotic convection regime will settle at high enough $Ra$ once the boundary layer have undergone a laminar to turbulent transition. To the best of our knowledge, no alternative interpretation of all these observations is proposed any longer.

Despite the good agreement between observations and the theoretical prediction, two important issues still remain open. The first concerns the precise nature of the regime observed at very high $Ra$. In particular, what is the degree of overlapping between this observed ``Ultimate Regime'' (following the naming introduced in 1997) and Kraichnan's prediction ? Beside the experimental difficulty of reaching very high $Ra$ in laboratory experiments, this comparison is delicate due to the ill-defined concept of laminar-to-turbulent transition in unsteady boundary layers, such as the ones present in turbulence convection (for example, see \cite{duPuits_JFM2007,Verdoold2008}) and  due to Kraichnan's renouncement to treat the ``join'' between his asymptotic regime and the so-called ``hard turbulence'' regime present at lower $Ra$.%, not to mention the large uncertainty in his numerical estimations that is stressed across Kraichnan's original paper.

The second important open issue is the experimental conditions for the triggering of this Ultimate Regime, which is observed in some experiments but not in all. Indeed, if we consider heat transfer measurements reaching at least $Ra=10^{13}$ and fulfilling the Boussinesq approximations, the litterature reports two sets of results in apparent contradiction : a clear transition is found in some\cite{Chavanne1997,RochePRE2001,Kenjeres2002,Niemela2003,RochePoF2005,GauthierETC11_2007} and not in others  \cite{WuTHESE,Ashkenazi1999,Niemela2000,Amati2005}. %\textit{Tertium quid},
Adding to the complexity of the present situation, two ``in-between'' results evidenced some features of transitions at very high $Ra$ but without increase in heat transfer. The first is a  simulation showing that the friction coefficient on the thermal plates departs from the typical scaling of laminar boundary layers above $Ra\simeq 10^{12}$, ``presumably marking the transition to turbulence''\cite{Amati2005} . The second is an experiment  which identifies two transitions for $Ra\simeq 10^{11}$ and $Ra\simeq 10^{13}$, based on statistical analysis of the temperature fluctuations in the flow\cite{Procaccia1991,WuTHESE}. We will come back on these observations, referred later as the ``Chicago experiment''.

%\footnote{To quote Kraichnan, "Apart from the uncertainties in the numerical coefficients, it is expected that the various laws of functional dependence summarized above can be accurate only when the qualifying inequalities are all large. [...]. In view of the inaccuracies inherent in the mixing length approach, we think it would be largely illusory to correct discrepancies of this kind by more careful treatment of the joins between the various asymptotic regions".}

The motivation of this paper is to 
answer the question\,: %how can one detect the transition to the Ultimate Regime from a local temperature measurements within the flow\,? In other words, 
%aside from its integral signature (enhanced global heat transfer) and its boundary layer signature (enhanced fluctuations of the temperature drop), 
What is the signature of the transition to the Ultimate Regime within the flow itself ?
 Such a signature has already been reported by Chavanne \textit{et al.}  \cite{Chavanne2001} but a finite-probe-size argument - proposed by Grossmann and Lohse (GL) to re-interpret the Chicago experiment  \cite{Grossmann1993} - should also apply for the fluctuations measurements reported by Chavanne \textit{et al.}.% Indeed, these later measurements were conducted with similar probes as in Chicago.
We note here that both of these two experiments were conducted using cryogenic helium as a convecting fluid in order to achieve very high $Ra$ in Boussinesq conditions.
This paper is organised in three sections. In the first, we present a systematic experimental study of the probe-size dependence of the measured temperature fluctuations, along with an analytical model. In the second section, we discuss two previous works related to transitions at very high $Ra$. On the one side, we find that a key hypothesis of GL model -on the scaling of the velocity boundary layer around the probe- is not satisfied, calling into question the conclusion of this theoretical work. On the other side, we find that the temperature signal recorded by Chavanne \textit{et al.} is altered by a finite-size correction, calling for a confirmation of their results with a probe having a space--time resolution at least 3 times better. The third section of this paper reports on the temperature fluctuations obtained with a specially made 17-$\mu$m thermometer, nearly 12 times smaller that the ones used previously. These measurements are backed-up by a systematic study conducted in the Barrel of Ilmenau.

\section{Finite size correction in local thermometry}

\subsection{Set-up}

\begin{figure}[ht]
\onefigure[width=\columnwidth]{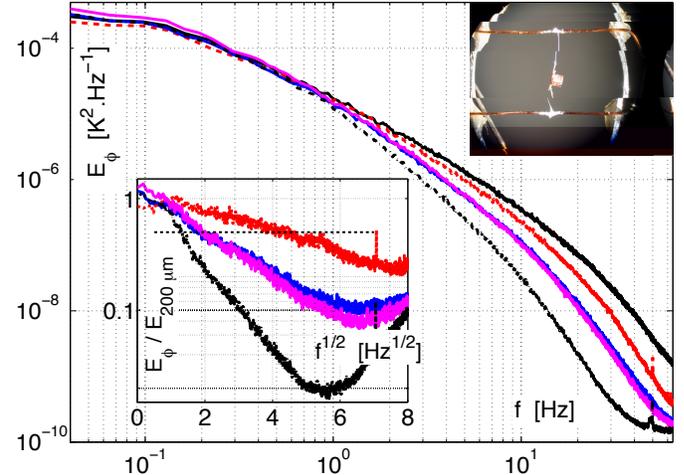}
\caption{Power spectral density $E_{\phi}$ of the temperature fluctuations sensed by probes of various sizes $\phi$ : from top to bottom (at 20 Hz), $\phi=$ 200 $\mu$m, 500 $\mu$m, 1 mm, 1 mm and 2 mm
(${Ra} = 2.0\times 10^{11}$). Lower insert : Spectral densities ratio $E_{\phi} / E_{200\mu m}$ versus $\sqrt{f}$. From top to bottom, $\phi=$500 $\mu$m, 1 mm, 1 mm and 2 mm. Upper insert : Top view of a 1-mm probe.
} \label{fig:SpectralDensity}
\end{figure}

Five thermometers were made with typical sizes $\phi$ ranging from $\phi$=200\,$\mu$m to 2 mm and aspect ratios close to 1. The 200-$\mu$m-probe was a cube of As-doped Si, similar to the ones used previously in Chicago and by Chavanne \textit{et al.}. Probes of sizes 500\,$\mu$m, 1 and 2 mm where assembled by tightly varnishing together $300 \mu m \times 300 \mu m \times 30 \mu m$ AsGa substrate thermometers \cite{Mitin2007} and cylinders of annealed copper of various sizes. The probes were contacted and suspended by two 50\,$\mu$m-diameter low-conductivity constantan wires -thermalised to the copper block- with length of several mm to minimize the intrusion of the support. The upper inset of Fig.\ref{fig:SpectralDensity} shows a probe of size 1 mm.
To check reproducibility of results, two probes of size 1 mm were made. 
The probes were calibrated and operated in the well-mixed core region of a 20 cm height and 10 cm diameter cryogenic Rayleigh-B\'enard cell, at equal distance from the cell axis and vertical side wall. %They were evenly spaced in a non-monotonic order versus size%\footnote{the sequence is 200\,$\mu$m, 1 mm, 500\,$\mu$m, 1 mm, 2 mm}
%.  The cell was filled with cryogenic He and the probes were calibrated in-situ.
 The unavoidable overheating due to the measuring current, corresponding here to a few hundreds of picowatt, was undetectable.

Fig.\ref{fig:SpectralDensity} shows typical temperature power spectra recorded by the different thermometers for  $Ra = 2.0\times 10^{11}$ ($Pr = 0.76$). We note a good collapse of the two spectra from the 1-mm thermometers %\footnote{The difference is compatible with a slight visible difference in size between these two probes.}
 as well as a reasonable superposition of spectra at low frequency. As expected, the larger thermometers have more pronounced roll-off at high frequency due to a larger space--time averaging.

Before presenting a model accounting for the finite-size effect, we recall the definition of the Rayleigh and Prandtl numbers\,:
\begin{equation}
Ra=\alpha \Delta g h^3 / \nu \kappa \mbox{ and } Pr= \nu / \kappa 
\end{equation}

where $\alpha$ is the isobaric thermal expansion coefficient, $\Delta$ the temperature drop across the cell, $h=20$\,cm the height of the cell, and $\nu$ and $\kappa$ are the kinematic viscosity and molecular thermal diffusivity of the fluid. The range of $Ra$ and $Pr$ explored with these probes spans from $4 \times 10^{10}$ to $2 \times 10^{14}$ and from $0.7$ to $7$ respectively.

The mean velocity of the large circulation in the convection cell was estimated from the cross-correlation of the temperature fluctuations seen by different probes. For example, a maximum of the cross-correlation between 2 opposite probes for a time delay of 6 s was interpreted as resulting for a large scale circulation with a typically velocity 20\,cm/6 s $\simeq$ 3.3\,cm/s.  These estimations were consistent with \textit{local} velocity measurements at mid-height by Chavanne \cite{Chavanne2001} in a similar cell. In this later work, the authors also derived a fit for the local velocity at mid height, and we used it to estimate the characteristic velocity $V$ seen by the thermometers.
%
%\begin{equation}
%Re = \frac{Vh}{\nu} \simeq 0.206 \times Ra^{0.49} \times {Pr}^{-0.7} \label{eq:ReFit}
%\end{equation}

\subsection{Modelization}

To account quantitatively for the finite resolution of a probe of size $\phi$, we define its transfer functions, $H_{\phi}(f)$, in Fourier space, as the magnitude of the measured temperature normalised by the temperature that would have measured a ideal probe (i.e. infinitely fast and small). 

The thermometers response involves several characteristic frequency scales. First, the frequency $f_{V}$ associated with the spatial filtering of the probe :
\begin{equation}
f_{V} = V/\phi
\label{eq:defFs}
\end{equation}
%$f_{V}^{-1}$ represents the advection time of a infinitesimal structure over the probe.

Second, the frequency $f_{\kappa}$ associated with the thermal response of the velocity boundary layer surrounding the probe\,:
\begin{equation}
f_{\kappa} ={\kappa} / {\lambda _{\phi}^2}
\label{eq:defFt}
\end{equation}
where $\lambda _{\phi}$ is the thickness of the probe's boundary layer  (defined quantitatively later). %$f_{\kappa}^{-1}$ represents the diffusion time across the boundary layer. 
Two other time scales associated with the probe are found to be significantly smaller than $f_{\kappa}^{-1}$ : the thermal diffusion time inside the probe and the ``$RC$'' time scale, where $R$ is the thermal resistance of the probe's boundary layer and $C$ is the heat capacity of the probe. This special hierarchy of times scales is often found in cryogenic hydrodynamics and results from the very low heat capacity of material at low temperature, and accordingly from their high thermal diffusivity.%At room temperature, the time scales can entail additional limitations on the time response of probes.

The velocity around the probe is time dependent and this reflects on the thickness of the velocity boundary layer with a typical viscous frequency response $f_{\nu} ={\nu} / {\lambda _{\phi}^2}$. This study is focused on fluids with intermediate Prandtl number ( $0.7<Pr<7$) implying that $f_{\nu}$ and $f_{\kappa}=f_{\nu}.Pr$ have the same order of magnitude. In the following, the dependence versus $f_{\nu}$ will be accounted implicitly via the dependence versus $f_{\kappa}$ and $Pr$. 

We now compare the frequency scales $f_{V}$ and $f_{\kappa}$ with experimental data in order to determine which causes most of the observed roll-off at high frequency.  Using decibels notations, we arbitrarily define the so-called -3 dB  cutoff frequency $f_{-3dB}$ of a probe of size $\phi$ by : 
\begin{equation}
H_{\phi}(f_{-3dB})=1/ {\sqrt{2}}
\end{equation}

A first order over-estimation of $f_{-3dB}$, called $f_{-3dB}^\prime$, is derived for the 3 larger probes using the approximated transfer function $H_{\phi}(f) \simeq \sqrt{ E_\phi(f) /E_{200\mu m}(f) }$, where $E_{\phi}$ is the power spectral density measured by the probe of size $\phi$.
The lower insert of Fig.\ref{fig:SpectralDensity} shows such spectral density ratio. 
The measured $f_{-3dB}^\prime$ can be compared with the -3 dB cut-off frequency expected from the sole spatial filtering associated with $f_{V}$. If this spatial filtering is modeled in 1 dimension by a moving average performed over a flat-top smoothing window of size $\phi$, the resulting transfer function will be $| {\sinc( 2\pi f \phi/V) }|$ and the -3 dB cut-off frequency will be $1.39 V / 2 \pi \phi \simeq 0.2 f_{V}$. This spatial cut-off is found to be 6 times larger -on average- than  the over-estimation $ f_{-3dB}^\prime$ (and at least 3 times larger in all cases), showing that the response of the thermometers is more limited by the diffusion time across the boundary layer rather than by the spatial resolution of the probes. Surely, this statement will no longer be true for asymptotically smaller probes.%\footnote{Using results derived later in this section to compare $f_{V}$ and $f_{\kappa}$, it is easy to show that spatial filtering becomes the limiting process for $Re_{\phi}Pr<10$ typically, where $Re_{\phi}$ is defined by Eq.\ref{eq:RePhi}.} since diffusion process scales like $\phi^{2}$ while the space filtering  scales like $\phi$.

The boundary layer thickness $\lambda _{\phi} $ is defined by comparing the measured transfer function with a generic transfer function for a diffusion-limited process :
\begin{equation}
H_{\phi} = e^{-{\lambda _{\phi}}/ \zeta } 
\label{eq:skin}
\end{equation}

\noindent
where $\zeta=\sqrt{ { \kappa} /{\pi f } }$ is thermal skin depth of the quiescent fluid.  The relevance of this analytical formula is supported by the reasonable linearity of the data plotted in the x-y representation chosen for the insert of Fig.\ref{fig:SpectralDensity}. If $f_{V}$ had been the most relevant frequency scale instead of $f_{\kappa}$, we would have expected transfer functions with a steeper roll-off. Using Eq. \ref{eq:skin} and the definition %s of $ f_{-3dB}$ and  
of $f_{-3dB}^\prime$, we get
%\footnote{$\lambda _{\phi}$ is estimated below from the -3dB cut-off frequency. Alternatively, it can be determined by fitting the whole transfer function with Eq.\ref{eq:skin}. This second approach was not favor because it turned out to required more arbitrariness when trying to superimpose the experimental spectra with the fit curve. Nevertheless, it gave results fully consistent with the one reported in the present work.}
 :
%\begin{equation}
%\lambda _{\phi} =  \beta / \sqrt{ f_{-3dB}}
%\label{eq:phif}
%\end{equation}
\begin{equation}
\lambda _{\phi} - \lambda _{200\mu m} = \beta / \sqrt{ f_{-3dB}^\prime}
\label{eq:phifp}
\end{equation}

%\noindent
%where 
%\begin{equation}
%\beta = \sqrt{{\kappa}/{4 \pi}} .{\ln 2}
%\end{equation}

%\begin{figure}[ht]
%\Image{rapport_acqui1}
%\caption{Ratio of the spectral energy of the bigger probes to the spectral energy of $T_2$} \label{figure:RapportSpectres}
%\end{figure}

%\begin{figure}[ht]
%\Image{Comparaison_Fc_fs}
%\caption{Comparison of the experimental cutoff frequency $f_{-3dB}$ and the spatial cutoff frequency $f_{V}$ for $T_{10b}$ ({\color{blue}$\blacktriangleleft$}), $T_{10}$ ({\color{magenta} $\blacktriangleright$}) and $T_{20}$ ({\color{green!50!black} $\blacklozenge$})} \label{figure:ComparaisonFcFs}
%\end{figure}

\noindent
where $\beta = \sqrt{{\kappa}/{4 \pi}} \cdot {\ln 2}$. Can we now fit $\lambda _{\phi}/\phi$, as a function of the Reynolds number $Re_{\phi}$ of the probe  ? 
\begin{equation}
{Re}_{\phi} = V  \phi/\nu
\label{eq:RePhi}
\end{equation}

In the ${Re}_{\phi} \gg 10$ limit, the thickness of a
boundary layer should be proportional to $\phi/\sqrt{{Re}_{\phi}}$ according to boundary layer literature\cite{Schlichting}. In the ${Re}_{\phi} \ll 10$, the thickness should be of order $\phi$. To accomodate the two limits, $\lambda _{\phi}$ is fitted by\footnote{
It is interesting to note the similarity between the proposed fit function Eq.\ref{eq:XcFit} for local thermometers and King's law for cylindral hot wires. According to King's law, the heat transfer from an overheated wire to a flow can be fitted by ${Nu} = a+b\sqrt{{Re}_{\phi}}$ where $a$ and $b$ are of order one (for $Pr\sim 1$), and $Nu$ is the Nusselt number of the hot wire% \cite{Welty}
. Writing that this $Nu$ is limited by the thermal resistance of a $\lambda _{\phi}$-thick boundary layer, we obtain a law similar to Eq.\ref{eq:XcFit}. 
} :
\begin{equation}
\frac{\phi}{\lambda _{\phi}} = A + B\sqrt{{Re}_{\phi}}{Pr}^{\alpha}
\label{eq:XcFit}
\end{equation}

\begin{figure}[ht]
\onefigure[width=\columnwidth]{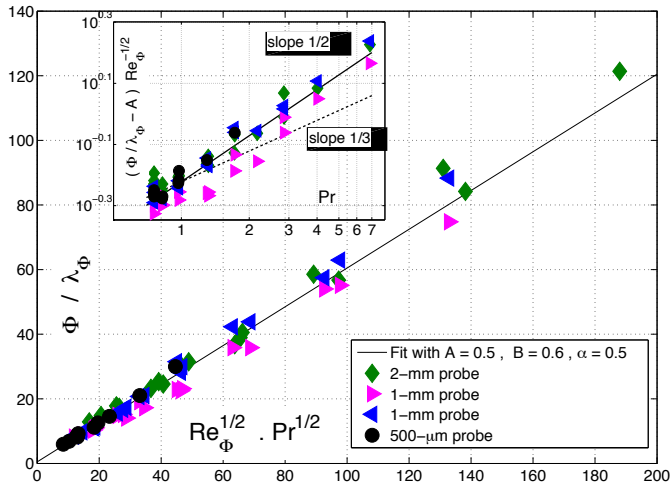}
\caption{Ratio between the size of the probe $\phi$ and the effective thickness of the probe's boundary layer, $\lambda _{\phi}$. The parameters $A$, $B$ and $\alpha$ are defined by Eq.\ref{eq:XcFit}. Insert :  $Pr$ dependence of this quantity compensated by $Re_{{\Phi}}^{1/2}$ (the offset $A= 0.5$ has a negligible contribution).%500 $\mu$m en noir,  $T_{10b}$ ({\color{blue}$\blacktriangleleft$}), $T_{10}$ ({\color{magenta}$\blacktriangleright$}) and $T_{20}$ ({\color{green!50!black}$\blacklozenge$}) )
}
\label{fig2}
\end{figure}

According to literature \cite{Schlichting}, $\alpha$ is expected to be $1/2$ for ${Pr} \ll 1$ and $1/3$ for ${Pr} \gg 1$. In the intermediate $Pr$ region of interest to us, we find that $Pr^{1/2}$ fits data slightly better than $Pr^{1/3}$, as shown on the insert of Fig.\ref{fig2}. Choosing for convenience this former $Pr$ dependence, all experimental data can be well fitted with the parameters $A=0.5$ and $B=0.6$, as illustrated by Fig.\ref{fig2}. The uncertainty on the value of $A$ could be as large as $0<A<2$ but we recall that this positive parameter of order 1 was just introduced to have a physically sound limit for $Re_{\phi} \rightarrow 0$. Thus, we retain as a fit validated up to $Re_{\phi} \simeq 6000$ and for $0.7<Pr<7$\,:
\begin{equation}
{\phi / \lambda _{\phi}} \simeq  { 0.5 + 0.6 \sqrt{{Re_{\phi} Pr}} }
\label{eq:LambdaFit}
\end{equation}

%\begin{figure}[ht]
%\onefigure[width=\columnwidth]{Encarts/X0XcEncarts}
%\caption{Ratio between the typical size of the probe, $\phi$ and the effective thickness of the boundary layer, $\lambda _{\phi}$, compensated by $\sqrt{{Re}_{\phi}}$ versus ${Pr}$ for $T_{10b}$ ({\color{blue}$\blacktriangleleft$}), $T_{10}$ ({\color{magenta}$\blacktriangleright$}) and $T_{20}$ ({\color{green!50!black}$\blacklozenge$}) and the resulting fits (solid line: $\alpha=1/3$, dashed line: $\alpha=1/2$)}
%\label{fig:PrvsRe}
%\end{figure}

Interestingly, from Eq.\ref{eq:LambdaFit} we find that the ratio $f_{V}/f_{\kappa}$ is constant ($\simeq 3$) at large $Re_{\phi}$ ($Re_{\phi}Pr \gtrsim 100$ typ.), showing that the spatial and time filtering scale similarly with $Re_{\phi}$. The exact value of this ratio results from arbitrary choices and definitions, but we showed previously that time filtering is the limiting process in our conditions. This hierarchy between the two filtering process could therefore be valid over a wider range of conditions that the ones explored in the present study.

As a summary, Eq.\ref{eq:skin} and Eq.\ref{eq:LambdaFit}, enable to predict the typical finite size-correction for a local thermometer, or compare quantitatively the time response of different probes. In the next section, we apply these results to two published studies on flow transitions at very high $Ra$.

\section{Consequences on the transition at high $Ra$}

\subsection{On Grossmann and Lohse's objection}

The transitions observed in Chicago \cite{Procaccia1991} for ${Ra} \simeq 10^{11}$
and ${Ra} \simeq 10^{13}$ have been re-interpreted as a probe-size artefact by Grossmann and Lohse (GL) \cite{Grossmann1993}.
In their model, GL first assume that the response of the probes is limited by thermal diffusion in the velocity boundary layer. This first assumption is in agreement with our findings.
A second assumption states that the probe's boundary layer thickness is  equal to the typical size of the probe for ${Ra} \leq 2\times 10^{13}$ and equal to the thickness $\lambda=h/2Nu$ of the thermal boundary layer over the plates of the convection cell for ${Ra} > 2\times 10^{13}$. For information,  the probe Reynolds number in the Chicago experiment reaches  $Re_{\phi} \simeq 400$  when ${Ra} = 2\times 10^{13}$ \cite{WuTHESE}.

%\begin{figure}[ht]
%\Image{Comparaison_Lohse}
%\caption{Comparison of the Grossmann and Lohse prediction ($\mathbf{-~-~-}$) with our measurements for bigger temperature probes, 
%$T_{10b}$ ({\color{blue} $\blacktriangleleft$}), 
%$T_{10}$ ({\color{magenta} $\blacktriangleright$}) 
%and $T_{20}$ ({\color{green!50!black} $\blacklozenge$}), 
%and with a fit for the Chicago experiment ({\color{red} $\circ$})} \label{fig:ComparaisonLohse}
%\end{figure}

The range of parameters $Ra$, $Pr$ and $Re_{\phi}$ explored in the Chicago experiment overlaps significantly with the one of the present study. Our analysis is therefore expected to hold. Our experimental results and model disagree with GL model in several ways. First,  for $Ra=2\times 10^{13}$, GL assumes that the probe's boundary layer thickness is comparable to the probe size $\phi$ while our analysis predicts a significantly smaller thickness\footnote{B. Castaing already pointed the surprising large boundary layer thickness assumed in the GL model during the \textit{Conference on High Rayleigh Number convection} at the ICTP in 2006.}, of order $\lambda \simeq  \phi/13$, due to the large $Re_{\phi} \gg 1$. Second, we find that the thickness of the probe's boundary layer is proportional to the probe size $\phi$. This differ from GL hypothesis of a boundary layer proportional to the convection cell height $h$ and independent of the probe size ${\phi}$ in the high $Ra$ limit.
Third, the $Ra$ dependence of the boundary layer thickness differs significantly.  GL model predicts two successive scalings : $Ra^0$ for  $Ra \leq 2 \times 10^{13}$, (corresponding to $Re_{\phi}  \lesssim 400$) and $Nu^{-1}\sim Ra^{-0.315\pm0.02}$ at higher $Ra$ while our measurements and model evidence a unique $Re_{\phi}^{-1/2} \sim Ra^{-1/4} $ scaling over nearly 4 decades of $Ra$ up to $Ra=2\times 10^{14} $ and for $Re_{\phi}$ spanning the range 77 to 6000 (see Fig.\ref{fig2} ) .

As a conclusion, our experimental results and analysis disagree with an important hypothesis of GL model.
Consequently,  the question about the possibility of a finite-size artefact in the measurements reported in Chicago \cite{Procaccia1991}, and consequently in those of Chavanne \textit{et al.} \cite{Chavanne2001} -of interest to us- remains fully open. This question is adressed  in the following sections.

\subsection{On Chavanne \textit{et al.} temperature fluctuations transition}

To test if the change in the temperature statistics reported in \cite{Chavanne2001} results from a finite-size artefact, one needs to compare the magnitude of the reported observation with the magnitude of the finite-size correction. Unfortunately, this comparison is delicate because the transfer function associated with the finite size correction is not fully known,  in particular its analytical dependence at small $f/f_{-3dB}$. Nevertheless, we can test if the magnitude of the observed transition is the same when measured with a $200 \mu m$ and $500 \mu m$ probes.

Following Chavanne \textit{et al.}, we consider the exponent $\xi_{2} $ of the 2$^{nd}$ order structure function of the temperature fluctuations $T(t)$\,:
\begin{equation}
\xi_{2}  = \frac{{d}\log{\left< \left( {T(t+\tau)-T(t) } \right) ^2 \right>_{t}}}{{d}\log {\tau} }
\end{equation}

\noindent
where the brackets represent time averaging. The dependence of $\xi_{2}$ versus the time increment $\tau$ (or versus space increment $V \times \tau$) contains the same mathematical information as a temperature power spectrum. On the Fig. 15 of ref.\cite{Chavanne2001}, the authors observed that the inflexion of  $\xi_{2}(V  .  \tau)$ differs below and above $Ra \simeq 10^{12}$, in particular for spatial increments in the window $0.5<V . \tau<2$\,cm. As shown on the insert of Fig.\ref{fig:D6Incr}, a similar qualitative change of shape is also present on our data, as illustrated with $Ra=3.6\times 10^{10}$ and $Ra=2.0\times 10^{12}$. But in the window of increments of interest, we also find that the measured $\xi_{2}$ depends of the size of the probe, here $200\,\mu$m versus $500\,\mu$m. This probe-size dependence reveals some finite size effects at least on the $500\,\mu$m probe, and up to increment $V . \tau \simeq 4$\,cm. 
To safely clear the increments window 0.5-2\,cm from visible finite size correction, the resolution of the 500 $\mu$m probe should be improved by a factor $4\,cm / 0.5\,cm=8$, corresponding to a reduction in size by a factor 8 according to the probe-size model. The 200 $\mu$m probe is therefore not small enough to be free from finite size effect in the range of increments of interest. A  probe of typical size $500/8 \simeq 60\,\mu$m is required.

As a conclusion, in the previous observations of Chavanne  \textit{et al.}, the interesting range of scales is not fully exempt of probe size effects. The signature of the transition could remain partly significant, but a confirmation with a much smaller probe would be desirable.

%As a conclusion, we cannot discard at this stage the possibility that the transition reported by Chavanne \textit{et al.} on the temperature fluctuations statistics is not an artefact. This motivated an additional experimental study presented in the following section.

%\begin{figure}[ht]
%\Image{X0XcRe05Pr05}
%\caption{Ratio of the typical probe size, $\phi$ and the effective thickness of the boundary layer, $\lambda _{\phi}$, as a function of $\sqrt{{Re}_{\phi}}\sqrt{{Pr}}$ for $T_{10b}$ ({\color{blue}$\blacktriangleleft$}), $T_{10}$ ({\color{magenta}$\blacktriangleright$}) and $T_{20}$ ({\color{green!50!black}$\blacklozenge$}) and the simpler fit (solid line)}\label{fig:FitSimple}
%\end{figure}

%\begin{figure}[ht]
%\Image{Re0vsxe1}
%\caption{$x_e/\phi$ computed from equation \ref{eq:xeFit} for ${Pr} = 1$. The horizontal bars represent the range of explored ${Re}_{\phi}$ for each experiment.} \label{fig:XeX0}
%\end{figure}

\section{Measurements with a micron-size thermometer}

To test directly the statistical signature of transition within the flow, we designed and micro-machined a thermometer which is one order of magnitude smaller than the ones used in the previous cryogenic convection experiments and 3.5 times smaller than the conservative $60$\,$\mu$m requirement defined above. The main characteristics of the probe are the following. The temperature sensitive area is a $\simeq 1 \mu$m large layer of annealed NbN \cite{Bourgeois:2006}  deposited across a 3 mm-long 17-$\mu m$-diameter  glass fiber. On each side of the temperature-sensitive area, a 150-nm-thick layer of gold is deposited along 1.5 mm of fiber, to provide electrical contacts with limited lateral heat conduction. The fiber is suspended at its ends by two 50 $\mu m$-diameter wires, themselves suspended on 220 $\mu m$ Cu wires glued across a thin ring of diameter 15 mm.  This web-like structure -already validated with micron-size hot wires \cite{Chanal1997,PietropintoPhysicaC2003}- was chosen to minimize the non-invasive character of the support %(see insert of Fig.\ref{fig:D6IncrInsert1})
. The temperature sensitivity was $\partial \ln R / \partial \ln T \simeq 0.7$ at 5K and the performance were limited by a resistance noise of spectral density $1.5 \times 10^{-6} /f$, corresponding to a rms noise of order 1mK. This noise prevented operation above $Ra=5 \times 10^{13}$ in order to maintain Boussinesq conditions. It also deterred us from putting the micro-thermometer in the middle of the cell as previously, due to the smaller temperature fluctuations here. The measuring current (0.5 pA) produced no detectable overheating\cite{SalortETC12_2009}. This micro-thermometer was suspended 2 mm above the bottom plate of a 43-cm-high 10-cm diameter convection cell. In the range of $Ra$ explored ($7\times 10^{10}<Ra<5 \times 10^{13}$), the estimated thermal boundary layer above the plate is always smaller than 1 mm and the probe can be considered to be outside of it. We will come back on this point later with a systematic study versus the plate-probe distance.

\begin{figure}[ht]
\onefigure[width=\columnwidth]{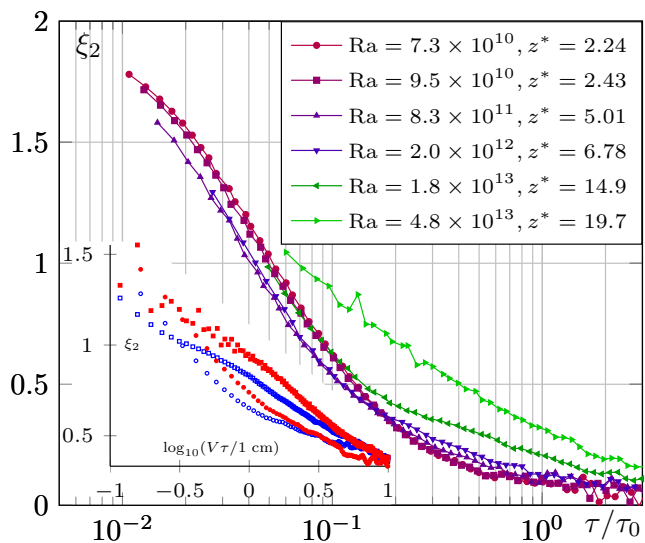}
\caption{Structure function exponent $\xi_{2}$  versus the time increment $\tau / \tau_{0}$ (see text) from the $17\,\mu$m thermometer.
For the 2 highest $Ra$, the shape of the curve changes. Insert : $\xi_{2}$ at  $Ra=3.6\times 10^{10}$ (circles) and $Ra=2.0\times 10^{12}$ (squares) from probes of size 200 (open symbols) and $500\,\mu$m (filled symbols). For direct comparison with the Fig.15 of ref. \cite{Chavanne2001}, the increments on the $x$ axis are in cm.}
\label{fig:D6Incr}
\end{figure}

On Fig.\ref{fig:D6Incr}, $\xi_{2}$ is plotted versus the dimensionless time increment $\tau/ \tau_{0}$ where $\tau_{0}= \lambda^2/\kappa$ is a caracteristic small time scale of the boundary layer. This normalisation of $\tau$ is chosen for convenience, the typical local velocity being unknown in this conditions. 
Changes in the spatial arrangements of the large-scale circulation inside the cell can induce offsets of the typically velocity seen by the probe and of the heat transfer $Nu$, up to a few tens of \%, which would result in a offset along the x-axis of this figure. Thus, the exact abscisa values result from an arbitrary normalisation and they are expected to be less universal than the general shape of the curve itself, which is robust to such offsets. A central result is that
-for $5\times 10^{-2}<\tau / \tau_{0} < 2$ typically- the shape of  $\xi_{2}(\tau/ \tau_{0})$ changes above ${Ra} \simeq 10^{13}$, corresponding also to the $Ra$ for which the heat transfer transition is found in this elongated cell\cite{SalortETC12_2009}.
The observed change in shape is qualitatively consistent with Chavanne \textit{et al.}'s \cite{Chavanne2001}, confirming a-posteriori that their observation was not an artefact although it may have been partly altered by a finite size effect.

\subsection{Effect of the distance plate-probe}

Effective distance between the micro-thermometer and the bottom plate in units of plate boundary layer thickness is $Ra$-dependent. One could object that the probe may not be fully outside the boundary layer till $Ra=10^{13}$ and that the observed transition corresponds to the exiting from some hypothetical outer boundary layer.

As a first comment, we note that such an artefact had no reason to appear for the same $Ra$ as the heat transfer enhancement, unless unlikely coincidental circonstances.

As a second comment, we consider the quantity $z^\star$ in the legend of Fig.\ref{fig:D6Incr}, which represents the remoteness of the probe from the plate in units of plate boundary layer thickness $\lambda=h/2Nu$.  We find that the estimated plate's boundary layer thickness represents less than 15\% of the plate-probe distance when the transition is observed. A-priori, when the transition is seen, the probe is already well outside the plate's thermal boundary layer. 

\begin{figure}[ht]
\onefigure[width=\columnwidth]{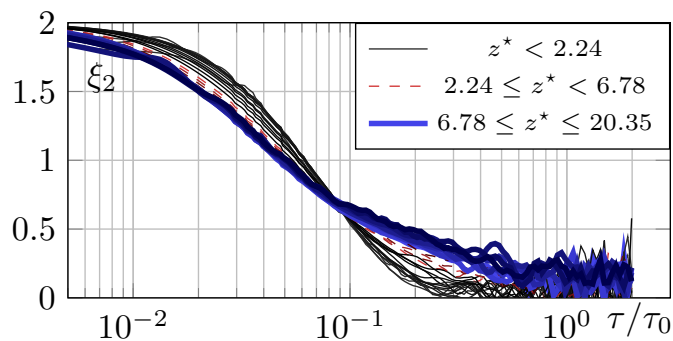}
\caption{
The exponent $\xi_{2}(\tau / \tau_{0})$ is plotted at different distance $z^{\star}=\frac{2Nu}{h}z$ from the bottom boundary layer for  ${Ra} = 3.2\times 10^{11}$ in the Barrel of Ilmenau.} \label{fig:IlmenauZ}
\end{figure}

To rule out directly the possibility of such an artefact, we carried a systematic analysis on the structure function exponent $\xi_{2}$ for different plate-probe distance. This study was carried in the ``Barrel of Ilmenau'', a Rayleigh B\'enard cylindrical cell filled with air (Pr=0.7) with an height set to 6.3 m, corresponding to an aspect ratio $\Phi / h$ of 1.13  \cite{duPuits_JFM2007}.
The local temperature fluctuations have been recorded using a temperature probe of size $150~{\mu m}$ that was moved along the central
axis of the cell above the bottom plate at $Ra=3.2 \times 10^{11}$. Fig.\ref{fig:IlmenauZ} illustrates that the $z^\star$-dependence of $\xi_{2}(\tau / \tau_{0})$ between $z^\star =6.78$ and $z^\star =20.35$ is significantly smaller than the magnitude of the change of $\xi_{2}(\tau / \tau_{0})$ observed between $z^\star =6.78$ and $z^\star =19.7$ in Fig.\ref{fig:D6Incr}.
Therefore, a plate-probe distance artefact can't explain the observed transition.%We note that the shape of $\xi_{2}$ only slowly evolves for $z^\star > XXX$, illustrating that the probe has entered the convection region often referred as the "bulk" or "core" of the flow.

\section{Concluding remarks}

Thanks to a systematic study of probe-size corrections, we showed that earlier measurements of temperature fluctuations in the core of a high Rayleigh number flow  \cite{Chavanne2001} were altered by finite-size effects, but not according to the scenario proposed in \cite{Grossmann1993}. An important conclusion from \cite{Chavanne2001}  was therefore questioned : is there really a visible qualitative change in the statistic of temperature fluctuations in the core of the flow when the flow enters the Ultimate Regime ? The main result of the present study is a ``yes'' answer to this question, thanks to a specially designed micro-thermometer.

What can we learn about the Ultimate Regime ? As shown in Fig.\ref{fig:D6Incr},  the transition is associated with an increase of $\xi_{2}$ for time increments $\tau / \tau_{0} \sim 1$, corresponding to the space increment $x=V.\tau =V \lambda^2 / \kappa$. The boundary layer thickness $\lambda$ corresponds to the balance of diffusive and advective transport. Outside of it -where our measurement are done- advection is stronger ($V > \kappa / \lambda$), implying $x=V.\tau > \lambda$. Thus, the transition manisfests itself on scales $x$ larger than the thermal boundary layer thickness $\lambda$, %(absence of correlation would entail $\xi_{2}=0$),
which could suggest an instability mechanism affecting extended pieces of boundary layers.

In addition to the enhancements of heat transfer\cite{Chavanne2001} and boundary layer fluctuations\cite{GauthierShotNoise:2008}, the Ultimate Regime is thus characterised by a third signature, but more work is needed to understand the basic mechanism of the instability occurring at very high Rayleigh numbers. Without better understanding of this \textit{terra nova}, we remain unable to extrapolate measurements and models of convection to flows of geophysical and astrophysical interests.

%As a side comment, we want to stress the non-intuitive strength of the finite-size correction. For example, visible correction for a 500 $\mu m$ probes are found for space increments as large as $V.\tau \simeq 3$\,cm, that is 60 times larger than the probe size.% This results from the accuracy of the diagnostic achievable with the structure function, but probably also from the mild roll-off of the transfer function associated with the finite size correction, which entails corrections down to lower frequencies than the spatial filtering.

%Although we have no precise quantitative model accounting for the change on the temperature statistics in the core of the flow, the occurrence of a change is consistent with existing results. Indeed, temperature fluctuations in the core of the flow are often viewed as detached pieces of the boundary layers, which cover the plates of the cell and concentrate most of the temperature gradient. A transition has been reported in these boundary layers \cite{GauthierShotNoise:2008}. Therefore, the change of statistics away from the boundary layer could simply reflect the change in the boundary layer.

\acknowledgments
We thank  B. H\'ebral, B. Chabaud, Y. Gagne, F. Chill\`a and more especially B. Castaing for illuminating discussions. The assistance of T. Fournier, P. Diribarne and P. Lachkar in thermometer development is gratefully acknowledged.
Financial support was provided by the R\'egion Rh\^one-Alpes (301491302), the Procope program (Deutscher Akademischer Austauschdienst D/0707571, Minist\`ere des Affaires Etrang\`eres 17858YD) and by the Deutsche Forschungsgemeinschaft (TH 497/22-1).

\bibliographystyle{eplbib}

%\bibliography{biblio_TURBU}

\end{document}